\documentstyle[preprint,aps,graphicx]{revtex}

\def\red{
}
\def\black{
}

\def\URLtilde{\lower0.2em\hbox{$\tilde{\phantom{a}}$}}
\def\mycomm#1{\hfill\break\strut\kern-3em{\red\tt ====> #1\black}\hfill\break}
\def\mycommNL#1{\strut\kern-3em{\tt ====> #1}\hfill\break}

\catcode`\@=11 
\def\lsim{\mathrel{\mathpalette\@versim<}}
\def\gsim{\mathrel{\mathpalette\@versim>}}
\def\@versim#1#2{\vcenter{\offinterlineskip
        \ialign{$\m@th#1\hfil##\hfil$\crcr#2\crcr\sim\crcr } }}
\catcode`\@=12 

\def\eqref#1{(\ref{#1})}

\makeatletter
\def\hlinewd#1{\noalign{\ifnum0=`}\fi
\hrule \@height #1 \futurelet \reserved@a\@xhline}
\def\hwhiteline{\noalign
{\ifnum0=`}\fi\hrule
\@height 0pt\vskip 1.0ex\futurelet \reserved@a\@xhline}
\makeatother
\def\gray{\special{ps: 0.40 setgray}}
\def\black{\special{ps: 0.0 setgray}}

\newcommand{\mydraft}{
\newcount\timecount
\newcount\hours \newcount\minutes  \newcount\temp \newcount\pmhours

\hours = \time
\divide\hours by 60
\temp = \hours
\multiply\temp by 60
\minutes = \time
\advance\minutes by -\temp
\def\hour{\the\hours}
\def\minute{\ifnum\minutes<10 0\the\minutes
    \else\the\minutes\fi}
\def\clock{
\ifnum\hours=0 12:\minute\ AM
\else\ifnum\hours<12 \hour:\minute\ AM
\else\ifnum\hours=12 12:\minute\ PM
    \else\ifnum\hours>12
     \pmhours=\hours
     \advance\pmhours by -12
     \the\pmhours:\minute\ PM
     \fi
    \fi
\fi
\fi
}
\def\fullclock{\hour:\minute}
\begin{centering}
\gray
\font\Hugett  =cmtt12 scaled\magstep4
\hbox{\Hugett Draft:\today,\clock}
\black
\end{centering}
\vskip -1.7cm
$\phantom{a}$
} 

\def\beq#1{\begin{equation} \label{#1}}
\def\eeq{\end{equation}}

\newskip\humongous \humongous=0pt plus 1000pt minus 1000pt

\newif\ifdtup

\begin{document}
{\tighten
 \preprint {\vbox{
  \hbox{$\phantom{aaa}$}
  \vskip-0.5cm
\hbox{TAUP 2798-05}
\hbox{WIS/06/05-FEB-DPP}
\hbox{ANL-HEP-PR-05-11}
}}

\title{Is observed direct CP violation 
in  $B_d \rightarrow K^+ \pi^-$
due to new physics? Check standard model
prediction of equal
violation in $B_s \rightarrow K^- \pi^+$} 
\author{Harry J. Lipkin\,\thanks{e-mail: \tt
ftlipkin@weizmann.ac.il} }
\address{ \vbox{\vskip 0.truecm}
Department of Particle Physics \\
Weizmann Institute of Science, Rehovot 76100, Israel \\
\vbox{\vskip 0.0truecm}
School of Physics and Astronomy \\
Raymond and Beverly Sackler Faculty of Exact Sciences \\
Tel Aviv University, Tel Aviv, Israel\\
and\\
High Energy Physics Division, Argonne National Laboratory \\
Argonne, IL 60439-4815, USA\\
}
\maketitle
\begin{abstract}%

The recently observed direct CP violation in $B_d \rightarrow K^+ \pi^-$ has
raised suggestions of possible new physics. A robust test of the  standard
model vs. new physics is its prediction of equal direct CP violation in  $B_s
\rightarrow K^- \pi^+$ decay.  CPT invariance requires the observed CP
violation to arise from the interference between the dominant  penguin
amplitude and another amplitude with a different weak phase and a different
strong phase.  The penguin contribution to $B_d \rightarrow K^+ \pi^-$ is known
to be reduced by a CKM factor in $B_s \rightarrow K^- \pi^+$. Thus the two
branching ratios are very different and  a different CP violation is expected.
But in the standard model a miracle occurs and the interfering  tree diagram is
enhanced by the same CKM factor that reduces the penguin to give the predicted
equality. This miracle is not expected in new physics; thus a search for and 
measurement of the predicted CP violation in  $B_s \rightarrow K^- \pi^+$ decay
is a sensitive test for a new physics contribution. A detailed analysis shows
this prediction  to be robust and insensitive to symmetry breaking effects and
possible additional contributions.

\end{abstract}

\vfill\eject 

\section{Introduction - Conditions for conclusive tests for new physics}

The recent discovery of direct CP violation in $B_d \rightarrow K^+ \pi^-$
decays has raised the question of whether this effect is described by the
standard model or is due to new physics beyond the standard model\cite{ellis}.
Unfortunately a quantitative standard model prediction for the CP violation is 
impossible because of its dependence upon strong phases which cannot be
calculated from QCD in the present state of the art.  

A general theorem from CPT invariance shows\cite{lipCPT} that direct CP
violation can occur only via the interference between two amplitudes which have
different weak phases and different strong phases. This  holds also for all
contributions from new physics beyond the standard model which conserve CPT.
Thus the experimental observation of direct CP violation in $B_d \rightarrow
K^+ \pi^-$ and the knowledge that the penguin amplitude is dominant for this
decay require that the decay amplitude must  contain at least one additional
amplitude with both weak and strong phases  different from those of the
penguin.  The question now arises whether this additional amplitude  is a
standard model amplitude or a new physics amplitude.

A natural check for this question is to examine other related decays. The
absence of CP violations found in  the charged decay $B^+ \rightarrow K^+
\pi^o$ immediately raised suggestions for new physics\cite{ellis}. However, the
reasons for relating the charged and neutral decays are not really serious.
Although only the spectator quark
is different, the CP violation can very different. 

A more serious and detailed investigation\cite{Groros5}
has recently been presented. Here we propose  checking specifically those other
decays where the standard model predicts an equal or related direct CP
violation and where the prediction  satisfies the following two conditions:

\begin{enumerate} 
\item If experiment agrees with the prediction it will be difficult
to find a new physics explanation. Thus new physics is ruled out for this
particular direct CP violation.
\item If experiment disagrees with the prediction it will be difficult to fix up
the standard model to explain the disagreement.
\end{enumerate}

This leads directly to the $B_s \rightarrow K^-\pi^+ $  decay, whose branching
ratio is much smaller than that for $B_d \rightarrow K^+ \pi^-$, and where the 
CP violation might be expected to be very different. Yet the standard model
predicts equal direct CP violation.  

We first note that 
the equality follows from a
``miracle" which occurs in the standard model and is not expected in common new
physics models; namely a
relation\cite{Gronau} between the CKM matrix elements, in which the tree diagram
contribution is enhanced by exactly the same factor that the dominant penguin
contribution is reduced. Thus altough the branching ratio for the $B_s$ decay 
which depends upon the dominant penguin contribution is reduced relative to the 
$B_d$ decay, the direct CP violation remans the same. 

This miracle is specifically relevant to the standard model and not expected if
the CP violation arises from interference between the penguin contribution and
a new physics contribution without the same dependence upon CKM matrix elements. 

If on the other hand the experiment disagrees strongly with the prediction we
note and will show below that the prediction depends upon minimum assumptions
whose validity can be carefully checked. It will be very difficult to ``fix" the
standard model to explain the disagreement.

Thus the experimental search for CP violation in $B_s \rightarrow K^-\pi^+ $
decay can provide convincing crucial information regarding the pesence or
absence of new physics in these decays. 
 
To put this argument on a firm foundation
we generalize the U-spin symmetry prediction\cite{Gronau,Groros} that the recently
observed direct CP violation in the $B_d \rightarrow K^+ \pi^-$ decay must be
matched by approximately equal direct CP violation in $B_s \rightarrow K^-
\pi^+ $ decay, even though the branching ratios can be very different. The
result for this particular decay to charge conjugate final states can be  
obtained by standard model arguments which do not require full SU(3) or U spin
symmetry, are nearly  independent of detailed models and require only
charge conjugation invariance for all final state rescattering.  

\section{Simplifications in $B_d \rightarrow K^+ \pi^-$ and 
$B_s \rightarrow K^-
\pi^+ $ Decays}

The particular $B_d \rightarrow K^+ \pi^-$ and  $B_s
\rightarrow K^- \pi^+ $ decays are much simpler than the other decays
considered\cite{Gronau} in the full U-spin multiplets. 

\begin{enumerate}

\item The final states are charge conjugate. All strong final state
rescattering and their relative phases remain related by the unbroken charge 
conjugation symmetry. 

\item The final states are isospin mixtures with a relative phase between the
two isospin amplitudes which is changed in an unknown manner by strong final
state rescattering. Unbroken charge conjugation invariance preserves the phase
relations between transitions to charge conjugate states.  SU(3) symmetry
breaking destroys phase relations between transitions to  U-spin rotated
states that are not charge conjugate; e.g. between $K\pi$, $\pi \pi$ and  $K
\bar K$ states.   

\item The spectator quark flavor cannot be changed in these decays with
one and only one quark of this flavor in both the initial and final states.
This eliminates all diagrams in which the spectator quark participates in the
weak vertex.

\end{enumerate}
 
These simplifictions enable a much more robust derivation of the standard model
prediction. Experimental violations will provide much more robust
indications of new physics than other previously cited indications for new
physics\cite{ellis} based on predictions which assume U-spin symmetry,
factorization or neglect of certain diagrams.

These simplifications are not present in the charged decay  $B^+
\rightarrow K^+ \pi^o$ where the final state can contain two $u$ quarks  which
have the  same flavor as the $u$ spectator quark. In this decay other diagrams
can occur  with participation of the spectator quark; e. g. the annihilation
diagram and the color-suppressed tree diagram. These both depend upon the same
CKM matrix elements as the color-favored tree diagram. Thus although they can
be small in comparison with the dominant penguin diagram, they can easily
combine with the smaller color-favored tree diagram to produce a total
amplitude proportional to the same CKM matrix factor as the tree diagram with a
very different strong phase and therefore a very different CP violation. 

\section{Simplifications from CKM properties and charge conjugation invariance}
   
General properties of the CKM matrix in the standard
model show\cite{Gronau}
that the amplitude for the $B_d \rightarrow K^+ \pi^-$ decay is the sum of two
amplitudes proportional
respectively to the products of CKM matrices $V^*_{ub}\cdot V_{us}$ and
$V^*_{cb}\cdot V_{cs}$. 

\beq{damp1}
A(B_d \rightarrow \pi^-K^+) = 
V^*_{ub}\cdot V_{us} \cdot T_d + V^*_{cb}\cdot V_{cs} \cdot P_d
\end{equation}
where $T_d$ and $P_d$ are two independent amplitudes labeled to correspond with
the tree and penguin amplitudes in the conventional description, but with
no dynamical assumptions. Eq. (\ref{damp1}) is identical to eq. (2) of 
ref \cite{Gronau}, with the amplitudes $A_u$ and $A_c$ of ref \cite{Gronau} 
replaced by $T_d$ and $P_d$. 
The corresponding charge conjugate amplitude is 

\beq{damp2}
A(\bar B_d \rightarrow \pi^+K^-) = 
V_{ub}\cdot V^*_{us}\cdot \bar T_d + V_{cb}\cdot V^*_{cs}\cdot \bar P_d 
\end{equation}
where $\bar T_d$ and $\bar P_d$ are two more independent amplitudes 

The direct CP violation observed is proportional to the
product Im($V^*_{ub}\cdot V_{us}\cdot V_{cb}\cdot V^*_{cs})$

Similarly, the amplitudes for the $B_s \rightarrow K^- \pi^+$ decay 
and the charge conjugate decay can be written

\beq{samp1}
A(B_s \rightarrow \pi^+ K^{-}) = 
V^*_{ub}\cdot V_{ud}\cdot T_s + V^*_{cb}\cdot V_{cd}\cdot P_s
\end{equation}

\beq{samp2}
A(\bar B_s \rightarrow \pi^-K^+) = 
 V_{ub}\cdot V^*_{ud}\cdot \bar T_s  + V^*_{cb}\cdot V_{cd}\cdot \bar P_s 
\end{equation}
where  $T_s$,  $P_s$, $\bar T_s$,  and $\bar P_s$ are all independent 
amplitudes. Our equations (\ref{damp1} - \ref{samp2}) differ from the
corresponding equations (4-7) of ref \cite{Gronau} by keeping all eight
amplitudes independent and not introducing the SU(3) symmetry assumptions of
ref \cite{Gronau}. 

 The direct CP violation hopefully to be observed is proportional to the
product 

\noindent Im($V^*_{ub}\cdot V_{ud}\cdot V_{cb}\cdot V^*_{cd})$.

Although the individual terms in the $B_d$ and $B_s$ decays are very different 
and the branching ratios for the $B_d \rightarrow K^+ \pi^-$ and 
$B_s \rightarrow K^- \pi^+$ decays are very different, Gronau has 
shown\cite{Gronau} that the two relevant products of CKM matrix elements 
satisfy the relation  

\beq{prodeq}
Im(V^*_{ub}\cdot V_{ud} \cdot V_{cb}\cdot V^*_{cd}) =
-Im(V^*_{ub}\cdot V_{us} \cdot V_{cb}\cdot V^*_{cs}) 
\end{equation} 

Since the strong interactions for the transition between the quark level and
the final hadron states are invariant under charge conjugation, and the final
states are charge conjugate, all relevant products of TP amplitudes can be
expected to be approximately  equal and the CP violation to be approximately
equal for the two transitions. The validity of this assumption of approximate
equality is discussed in detail below.

We now first show how this equal CP violation follows from the conventional
description in which the two terms are called penguin and tree diagrams and the
$B_d \rightarrow K^+ \pi^-$ and  $B_s \rightarrow K^- \pi^+$ decays are U-spin
mirrors related by SU(3). We then present a more general derivation in which
the detailed dynamics of the two terms are not needed, all diagrams
proportional to these two CKM factors are automatically included and full SU(3)
symmetry is not required.     

\section{The U-spin prediction with penguins and trees}

A large number of SU(3) symmetry relations between $B_d$ and $B_s$  decays to
charge conjugate final states\cite{pengsu3}  were obtained by extending the
SU(3) symmetry relations found by Gronau et al \cite{ROSGRO}. 
This can be seen at the quark level by noting
the quark couplings in the penguin and  tree
diagrams for  $B_d \rightarrow K^+ \pi^-$ and 
$B_s \rightarrow K^- \pi^+$ decays related by the $d \leftrightarrow s$
U-spin\cite{Uspin} Weyl reflection:

\beq{qpenguin}
 B_d (\bar b d) \rightarrow_{penguin} (\bar s d G)
\rightarrow_{strong} K^+ \pi^-
; ~ ~ ~ ~ ~ ~
B_s (\bar b s) \rightarrow_{penguin} (\bar d s G)
\rightarrow_{strong} K^- \pi^+
\end{equation}

\beq{qtree}
 B_d (\bar b d) \rightarrow_{tree} (u \bar s) (\bar u d)
\rightarrow_{strong} K^+ \pi^-
; ~ ~ ~ ~ ~ ~
B_s (\bar b s) \rightarrow_{tree}  (u \bar d) (\bar u s)
\rightarrow_{strong} K^- \pi^+
\end{equation}

Although the weak penguin and tree transitions from the initial state to the
intermediate quark state are very different for $B_d$ and $B_s$ decays, the 
subsequent strong hadronizations from the intermediate quark state to the final  
hadronic state are strong interactions approximately invariant under SU(3) 
and its U-spin subgroup and exactly invariant under charge conjugation. They 
are expected to be equal for the $B_d$ and $B_s$ transitions into  final states
which are both U-spin mirrors and charge conjugate.
The analysis of SU(3) relations in B decays has recently been
updated\cite{Gronau,Groros,grosnir} and applied to CP asymmetries in $B_d$  and
$B_s$ decays. However, the particular role of charge conjugate final states has
not been emphasized. 

The penguin and tree contributions to $B_d$ and $B_s$ decays are proportional
to very different CKM factors and have different strong interactions.   These
differences introduce unknown parameters in any analysis. Thus even though the
strong interactions are approximately invariant under SU(3) and its U-spin
subgroup and are exactly invariant under charge conjugation, the branching
ratios and decay rates for the $B_d$ and $B_s$ decays depend upon unknown
combinations of the different tree and penguin amplitudes.

However, Gronau's theorem  (\ref{prodeq}) shows\cite{Gronau} that the products
of the tree and penguin contributions for $B_d$ and $B_s$ decays relevant to
direct CP violation are approximately equal with opposite sign. Thus the direct
CP violation observed in $B_d \rightarrow \pi^- K^+$ is related to the as yet
unobserved CP violation in $B_s \rightarrow \pi^+ K^-$.

\section{Possible complications from other diagrams like charming penguins}

A very different approach often called ``charming penguins" suggests  
significant contributions from final state interactions 
which produce a $K\pi$ final state by strong rescattering from a $D^* \bar
D^*_s$ intermediate state\cite{charpen}. It is
difficult to obtain a reliable quantitative estimate  of these contributions
along with their sensitivity to U-spin breaking, in particular for the strong
phase which is crucial for CP violation. But these contributions can be
appreciable\cite{charpen}. The experimental branching ratio\cite{PDG} for 
$B_d  \rightarrow D^{*+}\bar D^*_s$ is a thousand times larger than the
branching ratio for $B_d \rightarrow K^+ \pi^-$.

\beq{ddstar}
BR (B_d  \rightarrow D^{*+}\bar D^*_s )=
(1.9 \pm 0.5)\%; ~ ~ ~ ~ ~ ~
BR ( B_d \rightarrow K^+ \pi^-)
(1.85 \pm 0.11) \times 10^{-5}
\end{equation}

Thus a very small rescattering of this large amplitude can have a serious
effect on the strong interaction phase of the $B\rightarrow K\pi$ penguin
amplitude. Our present treatment avoids any quantitative estimate of the 
detailed dynamics of ``charming penguins". The sum of all such contributions
which are tree diagrams producing a $c \bar c$ pair subsequently annihilated by
a strong final state interaction is called an ``effective penguin diagram" 
because its dependence on the CKM matrix elements is the same as that of the
normal  penguin.

\beq{qpengeffd}
 B_d (\bar b d) \rightarrow_{pengeff} (\bar c d) (c \bar s)
\rightarrow_{strong} \bar D^{*-} D^*_s \rightarrow_{strong} K^+ \pi^-
\end{equation}
\beq{qpengeffs}
B_s (\bar b s) \rightarrow_{pengeff} (\bar c s) (c \bar d) 
\rightarrow_{strong} D^{*+}\bar D^*_s \rightarrow_{strong} K^- \pi^+
\end{equation}

All our subsequent analysis holds when the contribution of this  ``effective
penguin" diagram is included. However other results for direct CP violation  
which depend upon U-spin relations between transitions to states which are not
charge conjugates can  suffer serious errors due to SU(3) symmetry breaking. A
symmetry breaking which produces effects of order 10 or 20 per cent in
branching ratios can produce large effects in relative strong phases which are
crucial for direct CP violations. 

In particular we note that the ``effective penguin" contribution 
to $B_d \rightarrow \pi^+ \pi^-$ of the 
U-spin analog of (\ref{qpengeffd})
\beq{qpengeffdd}
 B_d (\bar b d) \rightarrow_{pengeff} (\bar c d) (\bar c d)
\rightarrow_{strong} D^{*+}\bar D^{*-} \rightarrow_{strong} \pi^+ \pi^-
\end{equation}
is expected to be much less than in the case of
$ B_d \rightarrow K^+ \pi^-$.
The experimental branching ratio\cite{PDG}
for  $B_d  \rightarrow D^{*+}\bar D^{*-}$ is only 180 times larger than the
branching ratio for $B_d \rightarrow \pi^+ \pi^-$ instead of a thousand.
\beq{ddstarpi}
BR (B_d  \rightarrow D^{*+}\bar D^{*-} )=(8.7 \pm 1.8) \times 10^{-4}
; ~ ~ ~ ~ ~ ~
BR ( B_d \rightarrow \pi^+ \pi^-)
= (4.8 \pm 1.8) \times 10^{-6}
\end{equation}

\section{A general formulation with minimum assumptions}

We now present a general formulation with the minimum assumptions
necessary to predict the CP violation to be observed in $B_s \rightarrow \pi^-
K^+$.  

The following simplifying features of the $B_d \rightarrow \pi^- K^+$ 
and $B_s \rightarrow \pi^+ K^-$ decays enable relating these decays
without the  U-spin assumptions needed in the general case. 

\begin {enumerate}

\item The spectator flavor is conserved in the transition and cannot participate in
a weak transition which necessarily involves flavor change. Thus the weak
transition involves only a weak $b \rightarrow q_f U \bar U$ or 
$\bar b\rightarrow \bar q_f U \bar U$  decay, where $U$ denotes either u, c or t
and $q_f$ denotes $s$ for $B_d$ decays and $d$ for $B_s$ decays 

\item Each decay amplitude can be described by  two terms proportional to two
different products of CKM matrices.  This is a general result following from
the flavor properties of the three $b$ or $\bar b$ decays noted above and the
unitarity of the CKM matrix. This description is expressed formally by eqs.
(\ref{damp1} - \ref{samp2})
where the labels 
$T$ and $P$ by analogy to the tree and penguin labels in the
conventional description imply no assumption of tree or penguin dynamics.
Eqs. (\ref{damp1} - \ref{samp2}) include all possible additional amplitudes allowed by
the standard model including electroweak and charming penguins. 

\item Direct CP violation can be observed only if the squares of these
amplitudes contain a product of two CKM matrix elements with different weak
phases and different strong phases.  

\item Only four independent products of four CKM matrix elements
are relevant to direct CP violation.

\item Our knowledge of QCD does not yet enable calculating strong
phases; however, the experimental observation of direct CP violation in $B_d
\rightarrow \pi^- K^+$ decays provides the information that the two terms must
have different strong and weak phases.

\end {enumerate}

So far there are no additional assumptions beyond those in the standard model. 
We now list our other  basic assumptions:
\begin {enumerate}

\item The amplitudes for all these decays factorize into a weak transition
described by products of CKM matrices and a strong factor invariant
under charge conjugation. Thus for transitions to two charge conjugate final
states denoted by $f$ and $\bar f$

\beq{factord}
T_d(f) = \bar T_d(\bar f) \equiv T(f); ~ ~ ~ 
P_d(f) = \bar P_d(\bar f) \equiv P(f)
\end{equation}

\item The mass difference between $B_s$ and $B_d$ is negelected. Thus decays to
the same and to charge conjugate final states have the same energy for  
$B_s$ and $B_d$ decays and the same strong decay factors 
$T(f)$ and $P(f)$.

\beq{factors}
T_s(f) = \bar T_s(\bar f) =  T(f); ~ ~ ~ 
P_s(f) = \bar P_s(\bar f)  = P(f); ~ ~ ~ 
\end{equation}

\item We neglect some hopefully small other U-spin-breaking 
effects arising from the $B_s$-$B_d$ mass difference and the
difference between pion and kaon form factors. 

\begin{itemize}
\item The mass difference produces
intermediate quark states and final $K\pi$ states with slightly different
energies and momenta. We neglect this dependence except for a small phase space
correction.

\item All transitions involve the product of a pion form factor and a kaon form
factor. These form factors are all equal in the U-spin symmetry limit and 
differences arising from symmetry breaking have been analyzed\cite{Gronau}. 
The tree-penguin interference term relevant to direct CP violation is
proportional to the product of four form factors, one of which is a pointlike 
form factor of the meson created from a $q \bar q$ pair produced at
the weak vertex of the tree diagram and the other three are hadronic.
The dominant symmetry-breaking in these products between
$B_d$ and $B_s$ decays is in the difference between the products of 
a pointlike kaon and a hadronic pion form factor for $B_d$ 
decay and of a pointlike pion and a hadronic kaon form factor for $B_s$ decay.
We neglect this symmetry-breaking here, but note that the error
introduced is expected to be real and not change the relative phase of diagrams
which is crucial for CP violation. The error can also be estimated from simple
models or detemined from other experiments\cite{Gronau}.

\item These are the only assumptions slightly related to U-spin. No other
symmetry between pions and kaons is assumed.

 \end {itemize}

\item The CKM matrices satisfy\cite{Gronau} Gronau's theorem 
(\ref{prodeq})

\end {enumerate}

We can immediately conclude that the strong and weak relative phases of the two
terms in the  $B_d \rightarrow \pi^- K^+$ decay amplitude are equal to the
corresponding relative phases in the $B_s \rightarrow \pi^+ K^-$ decay
amplitude, even though the magnitudes of these amplitudes are very different.

We now calculate the direct CP violation explicitly. 
Because the CKM factors are different, the U-spin symmetry 
breaking by the  CKM matrices is different for the tree and penguin
contributions. Thus simple U-spin relations have not been 
obtained\cite{pengsu3,ROSGRO} for branching ratios of transitions where both
contributions are appreciable. 

The direct CP violation in charmless strange $B_d$ and $B_s$ decays
to charge conjugate final states is insensitive to these
problems\cite{Gronau}.  Direct CP violation is proportional to interference
terms which depend upon   the CKM matrix elements via the products related by
Gronau's theorem eq. (\ref{prodeq}). Thus the $B_d$ and $B_s$ CP
violations each depend upon a single CKM parameter, products  insensitive to
the ratio of the tree and penguin contributions and related by eq.
(\ref{prodeq}). The CP violations in $B_d$ and $B_s$ decays to states which are
charge conjugates and  U-spin mirrors  thus depend to a good approximation on
equal single parameters.

Squaring eqs. (\ref{damp1} - \ref{samp2}) and substituting eqs.
(\ref{factord} - \ref{factors}) give the direct CP violations for $B_s 
\rightarrow K^- \pi^+$ and  $B_d \rightarrow K^+ \pi^-$

\beq{CPd}
|A (B_d \rightarrow \pi^-K^+)|^2 -
|A (\bar B_d \rightarrow \pi^+ K^-)|^2 =
4\rm {Im}(V^*_{ub}\cdot V_{us}\cdot V_{cb}\cdot V^*_{cs})
\cdot\rm {Im}(T \cdot P^*)
\end{equation}

\beq{CPs}
|A(B_s \rightarrow \pi^+ K^-)|^2 - 
|A(\bar B_s \rightarrow \pi^- K^+)|^2 =
4\rm {Im}(V^*_{ub}\cdot V_{ud}\cdot V_{cb}\cdot V^*_{cd})
\cdot\rm {Im}(T \cdot P^*)
\end{equation}
Eqs. (\ref{CPd}) and (\ref{CPs}) satisfy the CPT 
constraint\cite{lipCPT} that the direct CP violation vanishes unless the
amplitude contains two contributions for which both the weak and strong phases
are different. 

Combining Gronau's equality (\ref{prodeq}) with 
eqs. (\ref{CPd}) and (\ref{CPs}) gives

\beq{dirCPds}
|A(B_s \rightarrow \pi^+ K^-)|^2 - 
|A(\bar B_s \rightarrow \pi^- K^+)|^2 =
|A (\bar B_d \rightarrow \pi^+ K^-)|^2 -
|A (B_d \rightarrow \pi^- K^+)|^2 
\end{equation}

Since the individual tree and penguin contributions to U-spin conjugate $B_d$
and $B_s$ decays are very different and their branching ratios and lifetimes are
different, the equality (\ref{dirCPds}) does not apply to the expressions 
$A_{CP}$ commonly used to express CP violation. Instead we have 
\beq{ACPds}
A_{CP}(B_s \rightarrow \pi^+ K^-)=
A_{CP}(\bar B_d \rightarrow \pi^+ K^-) \cdot 
{{BR(B_s \rightarrow \pi^+ K^-)}\over{  
BR (\bar B_d \rightarrow \pi^+ K^-)}} \cdot {{\tau (B_d)}\over{\tau (B_s)}}
\end{equation}
where $BR$ denotes branching ratio and $\tau$ denotes lifetime.

The same derivation applies to decays to any higher $K^*$ resonance and any 
nonstrange isovector resonance.  
\beq{ACPdsres}
A_{CP}(B_s \rightarrow \pi^{*+} K^{*-}) =
A_{CP}(\bar B_d \rightarrow \pi^{*+} K^{*-}) \cdot 
{{BR(B_s \rightarrow \pi^{*+} K^{*-})}\over{  
BR (\bar B_d \rightarrow \pi^{*+} K^{*-})}} \cdot {{\tau (B_d)}\over{\tau (B_s)}}
\end{equation}

Since CPT requires that the lifetimes and total widths of the $B_d$ and 
$\bar B_d$ must be equal, the observed direct CP violation  (\ref{CPd}) must
be compensated by an equal and opposite direct CP violation in other $B_d$
decays. Furthermore, since CPT requires direct CP violation to vanish in any
eigenstate of the strong S matrix\cite{lipCPT}, this compensation must occur in
the set of states connected to $\pi^+ K^-$ by strong rescattering. Since parity
is conserved in strong interactions this excludes all odd parity states. 

It is not clear whether this compensation is spread over a large number of
multiparticle states or is dominated by a few quasi-two-body states. It will be
interesting to check this experimentally. In the toy
model of ref. \cite{lipCPT} the compensation occurs in the $\pi^o \bar K^o $
state connected to $\pi^+ K^-$ by charge exchange scattering. The next low mass
allowed state is the vector-vector state.

\section*{Acknowledgements}

This research was supported in part by the U.S. Department of Energy, Division
of High Energy Physics, Contract W-31-109-ENG-38. It is a pleasure to thank
Michael Gronau, Yuval Grossman, Yosef Nir, Jonathan Rosner,  Frank Wuerthwein
and Zoltan Ligeti for discussions and comments.

%
\catcode`\@=11 
\def\references{
\ifpreprintsty \vskip 10ex
%
\hbox to\hsize{\hss \large \refname \hss }\else
\vskip 24pt \hrule width\hsize \relax \vskip 1.6cm \fi \list
{\@biblabel {\arabic {enumiv}}}
{\labelwidth \WidestRefLabelThusFar \labelsep 4pt \leftmargin \labelwidth
\advance \leftmargin \labelsep \ifdim \baselinestretch pt>1 pt
\parsep 4pt\relax \else \parsep 0pt\relax \fi \itemsep \parsep \usecounter
{enumiv}\let \p@enumiv \@empty \def \theenumiv {\arabic {enumiv}}}
\let \newblock \relax \sloppy
 \clubpenalty 4000\widowpenalty 4000 \sfcode `\.=1000\relax \ifpreprintsty
\else \small \fi}
\catcode`\@=12 

\end{document}